# Reconfigurable Ubiquitous Computing Infrastructure for Load Balancing

K. Sellami, L. Sellami, P. F. Tiako

*Abstract*—Ubiquitous computing helps make data and services available to users anytime and anywhere. This makes the cooperation of devices a crucial need. In return, such cooperation causes an overload of the devices and/or networks, resulting in network malfunction and suspension of its activities. Our goal in this paper is to propose an approach of devices reconfiguration in order to help to reduce the energy consumption in ubiquitous environments. The idea is that when high-energy consumption is detected, we proceed to a change in component distribution on the devices to reduce and/or balance the energy consumption. We also investigate the possibility to detect high-energy consumption of devices/network based on devices abilities. As a result, our idea realizes a reconfiguration of devices aimed at reducing the consumption of energy and/or load balancing in ubiquitous environments.

## I. INTRODUCTION

UBIQUITOUS computing is applicable to all areas, with capabilities to the context sensitivity, invisibility and mobility [1] [2]. These systems known as ubiquitous allow mobility. Furthermore, the main advantage of these systems is to be sensitive to new environments (geo-location, availability of communications networks, user identification etc.) In other words, the system should be present everywhere and in the real time to serve man; the subject of the information [3]. Hence, such system deals with ambient intelligence. This definition implies that the implementation of ambient intelligence requires the mobilization of technological competence to design services tailored to the needs of users [4], [5].

In a ubiquitous system, we have a set of objects that need to communicate and interact with their environment. Environment and objects must be able to detect changes of context and adapt their behavior to these changes. This self-adaptation requires communication and cooperation between different devices (sensors and actuators) generating a high energy consumption [6].

In this article, we propose a change in component distribution on the devices to reduce and/or balance the energy consumption. To reduce the energy consumption in ubiquitous environments, we propose an approach of devices reconfiguration. The idea is that the reconfiguration is carried out when a high-energy consumption is detected in the system; this is due to the cooperation of environmental objects interaction, which should have a mobile access and ensure availability of data and services when needed.

The rest of the paper is organized as follows: Section II presents a background of ubiquitous systems, reconfiguration and load balancing environment. Section III details our proposal. Conclusion and perspectives are described in Section IV.

## II. BACKGROUND

### A. Ubiquitous Systems

Ubiquitous Computing (Ubicomp, UC) describes the omnipresence of computing. This is the name given to the third generation of computer systems.

In the first generation, organizations and businesses had computer machines, so each computer is used by many people at once. In the second generation, it is the personal use of computers where each computer is dedicated to a person.

The UC is characterized by the growth of interconnected mobile devices, in the form of Smart Phone, PDA and any device with a pre-integrated computer. Therefore, a person owns and uses multiple computers (mobile device) [7].

UC was introduced by Mark Weiser in 1988 [8] in his vision of twenty-first century computing. According to Weiser, ubiquitous computing is the integration of computer tools into objects of everyday life (used at home and at work), this idea informs us of the omnipresent network that makes technologies invisible, inseparable and interrelated.

UC is based on a set of technologies (hardware and / or software) [1], which is the third era in the history of computing; in order to have mobile access to data and processing and to offer the best qualities of service, the user will be at the center of the ubiquitous environment [4], [9]. The user has several available devices. These devices need to communicate and interact with their environment in order to be able to cooperate and to access necessary information. Users then have the possibility to easily exchange data, quickly and effortlessly, regardless of their geographic locations. This ubiquity way to access to information has a strong impact on society and it is also changing work habits when preserving privacy.

### B. Reconfiguration

The architecture describes how the processing, storage, and control elements are arranged and how they communicate. It is therefore both the description of the calculation scheme and

K. Sellami is with the LMA Laboratory, A/Mira University of Bejaia, Algeria (e-mail: skhaled36@yahoo.fr).

L. Sellami is with the Department of Computer Science, A/Mira University of Bejaia, Algeria (corresponding author; phone: 00213657348254; e-mail: slynda1@yahoo.fr:).

T. PF. Tiako is with Center of IT Research Development, Tiako University, Oklahoma, USA (e-mail: pftiako@tiakouniversity.org).

the communication plan.

The configuration describes the potential for an element to have several possible and distinct states of operation. Reconfiguration describes the possibility for an element to change state over time by a deterministic process. Indeed the change of configuration must be due to a specific order of reconfiguration and it must be possible to return to a previous state.

The system is consist of components (subsystems) some of which are complex system. the reconfigured system is the fact that behavior of the main system would be modified by aggregate effect of the various subsystem.

The reconfigurable system presents components pthat rovide some service in some configuration.

The architecture of the reconfiguration infrastructure consistst of component interfaces that facilitates understanding of its structure.

When interested in on-chip architectures we can consider two approaches to reconfiguration are considered [10]. A first approach focuses on the behavior of the system (application). When a modification of the system (application) is detected the reconfiguration mechanisms are set up in place in order to solve the problem by applying workflow adaptation actions to the existing architecture [11].
The second approach is to observe the components of the architecture when running an application. A reconfiguration of the architecture is brought when a change is detected.

TABLE 1 RECONFIGURATION APPROACHES "SELF-HEALING"

| Approach | Surveillance | Diagnostic | Reconfiguration action |
|---|---|---|---|
| Behavioral | Workflow level | An alarm leads eventually a reconfiguration. | Adaptation of the application to the architecture. |
| Architectural | At the system architecture level | A reconfiguration is considered after several symptoms | Applied to architecture. |

Reconfigurable architectures have become inevitable. They offer an alternative between the flexibility of programmable solutions and performance of specific circuits [12] [13].

New application fields are opened in order to be consistent with the properties of these architectures.

Reconfigurable architectures have changed the design of systems taking advantagev of their reconfiguration and adaptability characteristics.

Architecture can be described as reconfigurable if at least one of the elements it describes is reconfigurable, such as processing or communication resources.

We have two types of reconfiguration:

(1) Static reconfiguration: Consists in change a system when it is stopped. This may involve certain risks, such as obtaining an incoherent system. This is capable of causing disasters.

(2) Dynamic reconfiguration: Returns to the modification of a system while it is running. It applies mainly to systems that cannot be stopped in order to make a change. It thus preserves some system availability by reducing the downtime. Dynamic reconfiguration increases the efficiency of a system by allocating its logical resources to multiple tasks. Thus, by using the notion of virtual memory that allows to execute a larger program than the available physical memory, a dynamically reconfigurable system can offer a large amount of virtual logical resources and map them at run time to a reduced amount of physically available resources

*C. Load Balancing*

In computing, load balancing is a set of techniques to distribute workload among several computers in a group. These techniques allow both to respond to an excessive load of a service by distributing them to multiple servers, and reduce the potential unavailability of services that could cause the software or hardware failure of a single server.

Load balancing aims to optimize resources used, maximize throughput, minimize response time, and avoid overload of any single resource [14].

## III. THE PROPOSED WORK

OUR goal is to take a reconfiguration of devices aimed at reducing the consumption of energy in ubiquitous environments.
The method consists of a monitoring approach, interpretation, analysis and reconfiguration of ubiquitous systems running.

*A. Description of Proposition*

Our contribution is based on the principle of an ubiquitous network, which is composed of nodes (devices, sensors, hosts, etc.) [15].

Each node communicates with its neighbors and it has the knowledge of its environments (its neighbors). The network is divided into subset of nodes (cluster).

At the level of each cluster head (fig 1) is installed a controller responsible for analysis, detection and correction, high energy consumption detectection process.

For every membership of node in a cluster, the controller sends a DetectionAgent to the user node. The mission of the DetectionAgent is :

(1) To collect the information (data) related to the functioning and the level of energy consumption in the system.

(2) To analyze and compare the performance of the system.

(3) In the case of a malfunction possibly a change and / or overload of the device, the Controller specifies the appropriate actions to take to get out of the malfunction state and reconfiguration the system.

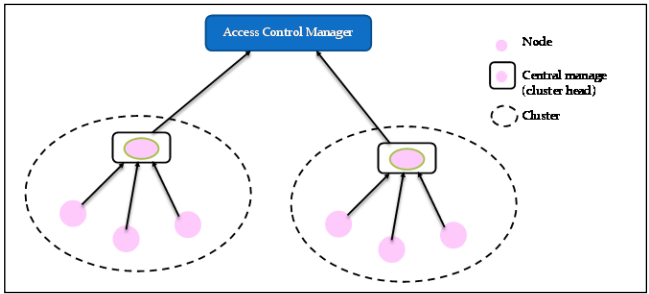

Fig. 1 SMART services architecture [16]

B. *Steps Detection of Energy Overload*

Figure 2 illustrates the functioning of our proposal; it consists of the following four parts: namely, Knowledge Base, Behavior, Control, and Action.

Knowledge Base is built based on the information (data) related to system operations, and the energy consumption of the device. This base is achieved.

Current Behavior of the user node is built by capturing information (data).

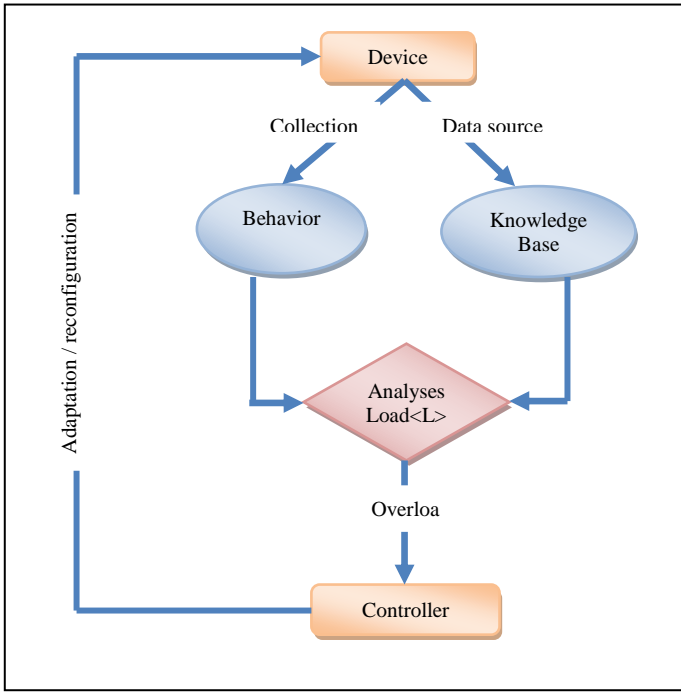

Fig. 2 Step of the proposal

To detect anomalies in the behavior node/energy overload, an analysis and Control is performed to compare the current behavior with the knowledge base of the device.

When detecting high-energy consumption, we proceed (Action) to a change in the distribution of the components on the devices, to reduce and / or balance energy consumption.

Our solution is essentially based on three components which are: Device, Detection Agent and Controller.

1) *Device:* The devices have a direct link with the user for the purpose of making personal profits. The device is designed to perform a task (eg: calculation). When a change of application (process / task) is made, the new task consumes energy overload on the device.
2) *Detection Agent:* Detection Agent collects data and compares them with the device knowledge base. When the current behavior of the device differs from its usual behavior, DetectionAgent informs the head of the controlling cluster.
3) *Controller:* The controller conducts adaptation to the new situation (application) of the device/system.

The controller undertakes aptitude measures which are:

(1) Sharing processing (calculation) ressources between multiple devices cooperating in the same cluster.

(2) Adapting application to the new situations : reconfiguring the device by improving its ability to reduce / equilibrate the load of energy consumption

## IV. EXPERIMENTAL RESULTS

TO simulate our approach, we used Georgia Tech Network Simulator (GTNet) [17] environment to prove the efficacity and applicability of our approach. The nodes used wireless connection to connect and communicate.

The services implemented in our simulation have limited capabilities (cf. Tab 2). To test the results of our approach, we overload the services to measure the system ability to detect overloading and make necessary correction.

TABLE 2
CAPACITY OF SERVICES- EXAMPLE

| Services | Print | View | Send e-mail | Update the BDD | Scanner |
|---|---|---|---|---|---|
| Normal | 34 | 123 | 10 | 50 | 8 |

Table 3 shows the detection results.

TABLE 3
DETECTION RESULTS

| Services | Print | View | Send e-mail | Update the BDD | Scanner |
|---|---|---|---|---|---|
| Overload | 50 | 124 | 21 | 56 | 30 |
| Detection | 50 | 124 | 21 | 56 | 30 |

In Table 3, we note that all overloading of services are detected, which proves the effectiveness of our proposal.

The Figure 3 shows the correction results.

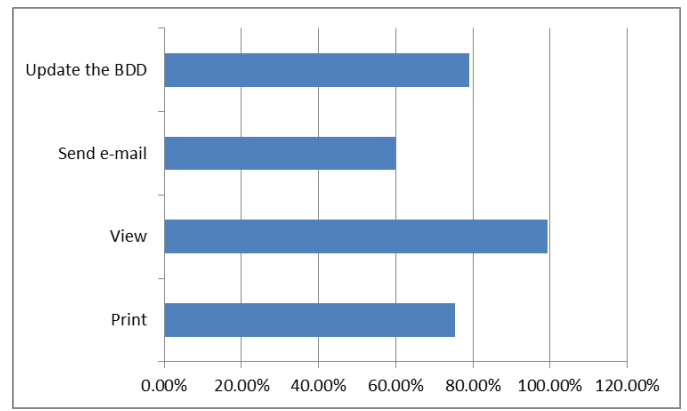

Fig. 3 Correction results

Figure 3 shows that in most cases, our proposal gives good results by correcting most errors (overloads).

## V. CONCLUSION

THE principal objective of this work was a reconfiguration and load balancing of devices aimed at reducing the consumption of energy in ubiquitous environments. We developed an approach to detect high-energy consumption. Our approach allows to monitor the energy consumption of nodes and the network. It consists in searching anomalies that could lead to possible overload, and to take action against such situation.

We described our approach focusing on the behavior of the node in the network.

This article is a work in progress presenting ideas to be developed in our future work.

In the future, we will complete the overload detection architecture, develop the method of reconfiguration, formulate and evaluate our approach of reduce the energy consumption.